\documentstyle[prb,tighten,twocolumn,epsfig,floats,aps]{revtex}

\def\fp{
\begin{figure}
  \psfig{file=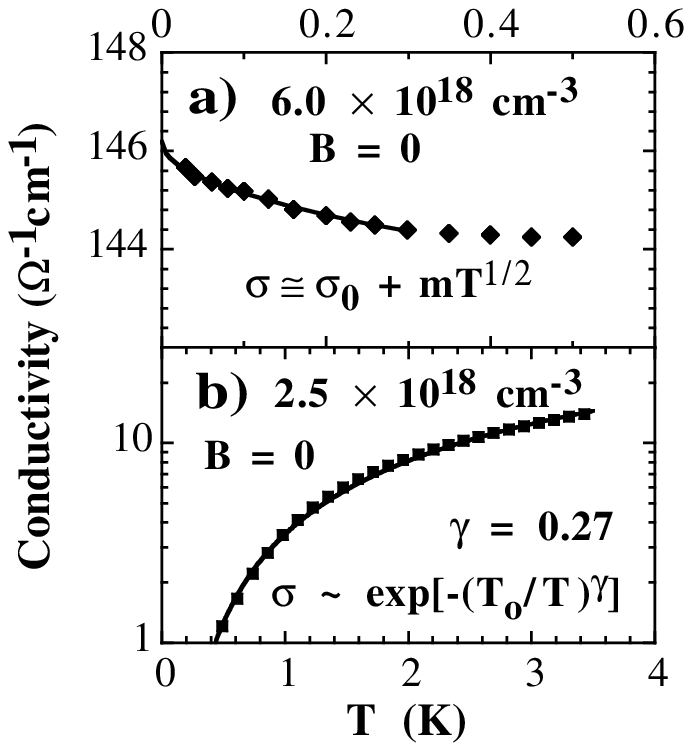,width=0.9\linewidth,angle=0}  
\caption{In-plane conductivity of Si/SiGe:Sb superlattices with
(a) $N = 6 \times 10^{18}$ cm$^{-3}$, and (b) $2.5 \times
10^{18}$ cm$^{-3}$ as a function of temperature in zero magnetic
field.  Solid lines represent fitting to formulae corresponding
to metallic (a) and hopping conduction (b), respectively.}
\label{fig:1}
\end{figure}
}

\def\fd{
\begin{figure}
  \psfig{file=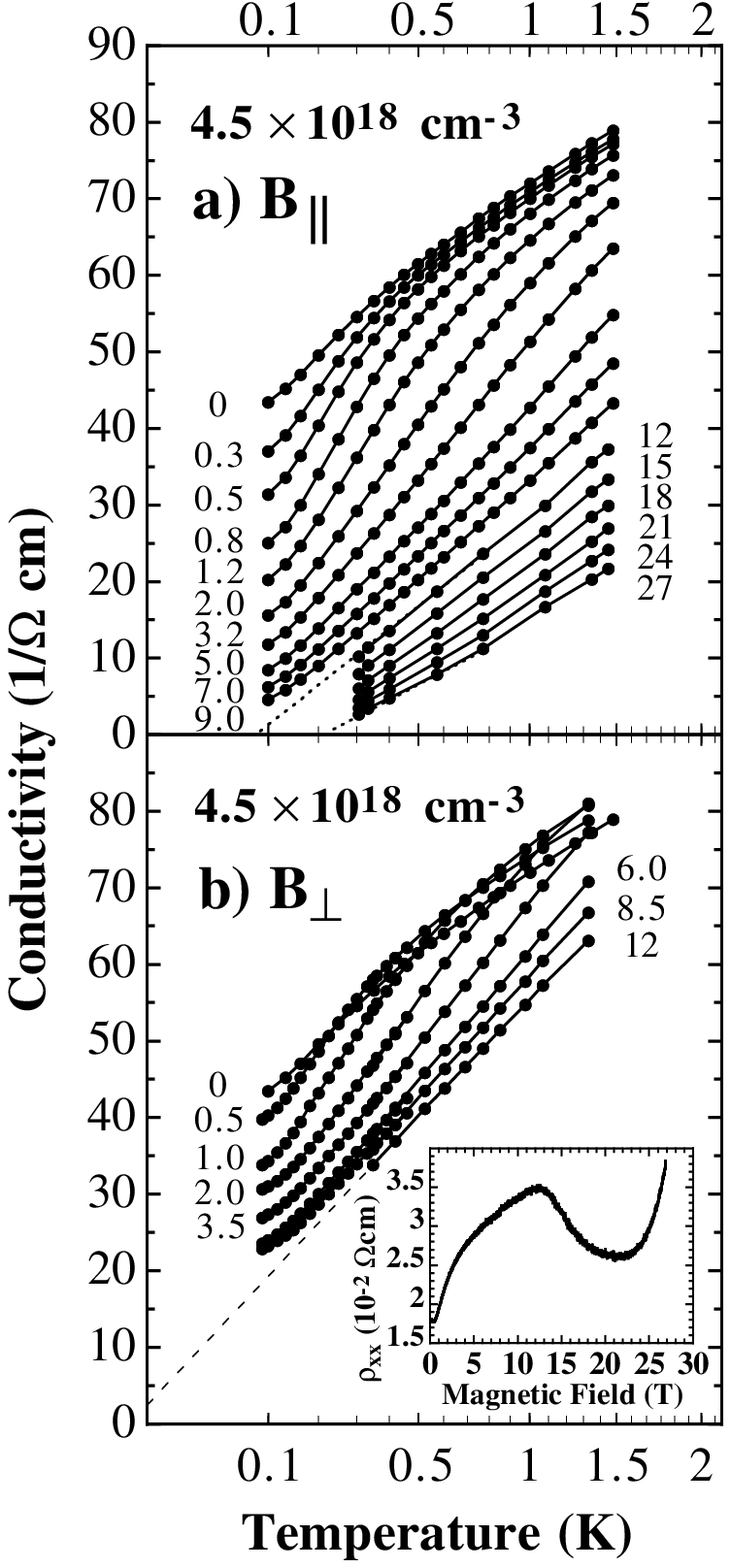,width=0.9\linewidth,angle=0}
\caption{Conductivity of Si/SiGe:Sb superlattice
with $\tilde{N} = 4.5$ in various magnetic fields (a) parallel and (b)
perpendicular to the layers as a function of temperature in $T^{1/2}$
scale.  The dashed lines are guides for the eye.  The inset in (b)
shows the magnetic-field dependence of the resistivity for the
perpendicular orientation.}
\label{fig:2}
\end{figure}
}

\def\ft{
\begin{figure}
  \psfig{file=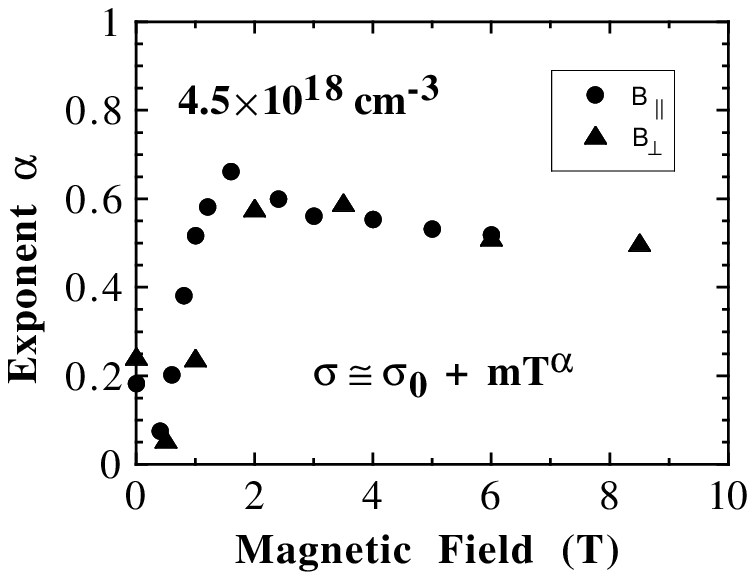,width=1\linewidth,angle=0}
\caption{Exponent $\alpha$ in the temperature dependence of conductivity
$\delta\sigma(T) \propto T^{\alpha}$ for parallel (circles) and
perpendicular (triangles) magnetic fields. }
\label{fig:3}
\end{figure}
}

\begin{document}

\title{Interaction effects at the magnetic-field induced 
metal-insulator transition in Si/SiGe superlattices}

\author{G. Brunthaler,$^a$ T. Dietl,$^{a,b}$ A. Prinz,$^{a,b}$
M. Sawicki,$^{a,b}$ J. Jaroszy\'nski,$^{b}$, P. G{\l}\'od,$^{b}$
F. Sch\"affler,$^{a}$ G. Bauer,$^{a}$\\ 
D.K. Maude,$^c$ and J.C. Portal,$^c$}
\address{
$^a$ Institut f\"ur Halbleiterphysik, Johannes Kepler
Universit\"at Linz, A-4040 Linz, Austria\\
$^b$ Institute of Physics, Polish Academy of Sciences,
al.~Lotnik\'ow 32/46, PL-02668 Warszawa, Poland\\
$^c$ High Magnetic Field Laboratory -- MPI-CNRS,
F-38042 Grenoble, France \\}

\maketitle

\begin{abstract}
A metal-insulator transition was induced by in-plane magnetic
fields up to 27 T in homogeneously Sb-doped Si/SiGe superlattice
structures.  The localisation is not observed for perpendicular
magnetic fields.  A comparison with magnetoconductivity
investigations in the weakly localised regime shows that the
delocalising effect originates from the interaction-induced
spin-triplet term in the particle-hole diffusion channel.  It is
expected that this term, possibly together with the singlet
particle-particle contribution, is of general importance in
disordered n-type Si bulk and heterostructures.
\\ 
\end{abstract}

Due to the sensitivity to external perturbations and the
possibility of controlling the randomness, doped semiconductors
and semiconductor structures have proven to be excellent media
to examine the nature of localisation in disordered electronic
systems.\cite{Alts85,Been91,Beli94} In particular, the
remarkable studies \cite{Rose83} of bulk Si:P have demonstrated
that the doping--induced metal--insulator transition is
continuous, and that the effects of backscattering and
disorder--modified electron--electron interaction are equally
important at the localisation boundary. The latter was confirmed
by investigations \cite{Uren80} of interfacial electron layers
in Si MOS-FETs, which indicate that all states are localised in
disordered two--dimensional (2D) systems. \cite{Abra79}  While
the corresponding theoretical predictions have found detailed
quantitative verifications in a number of experimental studies
on various semiconductor compounds and disordered metals, the
case of n-Si, \cite{Stup93} and more recently of high mobility
Si MOS-FETs, \cite{Krav95} have demonstrated
unsatisfactory understanding of the Si-based systems.  Extensive
theoretical studies \cite{Beli94} re-emphasise the peculiar
properties of n-Si, where the weakness of spin-dependent
scattering -- associated with the lightness of the constituting
atoms, the presence of inversion symmetry in the diamond
lattice, and the low degree of compensation -- makes the
anti-localising interaction terms in the conductivity
corrections particularly large.

Even more surprising are perhaps the results by Kravchenko et
al., \cite{Krav95} which point to the existence of a
metallic phase in high electron-mobility Si MOS-FETs at zero
magnetic field.  Several suggestions on the origin of this
unforeseen metallic phase in that 2D system have been put
forward: (i) Coulomb interactions of remote electrons,
\cite{Piku96} (ii) superconductivity due to spin-triplet
pairing, \cite{Beli97} (iii) non-Fermi liquid behaviour driven
by effects of electron-electron interactions, \cite{Dobr97} and
(iv) the existence of a spin-orbit gap due to a strongly
asymmetric confining potential in Si-MOS structures.
\cite{Puda97b}  In most recent experimental works
\cite{Simo97,Puda97} the suppression of the low temperature
conducting phase by an in-plane magnetic field was demonstrated,
which indicates the importance of spin-related terms.

Here, we report about millikelvin studies of in-plane
conductivity up to magnetic fields of 27 T  for uniformly
Sb-doped Si/SiGe superlattices (SLs).  In our Si/SiGe SLs, the
six fold valley degeneracy of the Si conduction band is removed
due to the biaxial strain. For the two ground state valleys the
in-plane effective mass is nearly by a factor of 5 smaller than
the longitudinal mass.  The vertical transport mobility is in
addition strongly reduced by the weak coupling through the SiGe
barriers, resulting in a rather large anisotropy of electron
transport.  At the same time, the electron Land\'e factor is
expected to be isotropic and close to its value in bulk n-Si,
$g^*=2.0$.  Accordingly, the magnetoresistance studies, which we
carried out for the two orientations of the magnetic field with
respect to the SL interfaces, make it possible to determine
the relative importance of the orbital and spin effects. Our
results confirm that electron delocalisation in Si-based
materials at the metal-insulator transition (MIT) is driven by
spin-dependent phenomena. We assign these phenomena to the
presence of anti-localisation terms in the quantum corrections
to the conductivity of disordered systems. In the absence of
spin-orbit coupling, such spin-dependent terms may arise from
the triplet particle-hole and singlet particle-particle
diffusion channels of the disorder-modified electron-electron
interactions. \cite{Alts85,Fink84}
  
{\fp}

The SL structures were grown by MBE on (001) Si substrates with
a SiGe graded buffer layer.  Each of the SLs consist of 150
periods of Si wells ($d_w = 25$ {\AA}) and
Si$_{0.55}$Ge$_{0.45}$ barriers ($d_b = 14$ {\AA}). The impurity
densities in our samples, $\tilde{N} \equiv N/10^{18}$ cm$^{-3}
= 2.5$, 4.5, and 6, cover the range around the critical
concentration of $\tilde{N} = 3$ for the MIT in bulk Si:Sb.
\cite{Long84}  The structural properties of the SLs were
determined by high resolution x-ray diffraction, whereas the Ge
and Sb concentrations and their homogeneity were checked by
secondary ion mass spectroscopy.  Resistivity and Hall effect
measurements between 10 and 300 K show only a small change of
the Hall constant with temperature, and lead to an agreement
between electron densities and Sb doping levels within 5\%.

Previous investigations of our structures, carried out in the
magnetic fields up to 9 T, have demonstrated that electron
conductivity retains features of 3D transport, despite that the
miniband width $w$ is more than 10 times smaller than the
in-plane momentum relaxation rate $\hbar/\tau$. \cite{Brun96} In
particular, the magnetoconductivity at 1.5 K in the weakly localised
regime can be described by the theory for a disordered 3D
electronic system with an anisotropic diffusion tensor.
\cite{Alts85,Alts81,Bhat85}

The present magnetoresistance studies were carried out in a
dilution refrigerator, installed either in a superconductive 9 T
coil or in a hybrid magnet capable of 27 T.  It has been found
in the course of this work that the temperature (and phase
coherence) of the conducting electrons in the studied structures
are strongly affected by high frequency electromagnetic noise.
It appears that the electrons in Si/SiGe are less resistant to
electron heating than in zinc-blend compounds. In the latter,
the piezoelectric coupling, due to its energy loss rate which is
proportional to $T_{e}^{3} - T_{l}^{3}$, results in better
electron cooling than that in the elemental semiconductors, in
which only the interaction by deformation potential,
characterised by $T_{e}^{5} - T_{l}^{5}$-dependence, is
effective at low temperatures. \cite{Ridl91} Here $T_{e}$
denotes the electron and $T_{l}$ the lattice temperature.

In Fig. 1, the temperature dependence of the in-plane
conductivity is shown for two Si/SiGe SLs at zero magnetic
field. For the highest doping level ($\tilde{N} = 6$) an
increasing conductivity with decreasing temperature is observed
(Fig. 1a). This metallic behaviour is well described by the
dependence $\sigma(T) = \sigma_{0} + mT^{1/2}$, where $m < 0$.
The lowest doped SL with $\tilde{N} = 2.5$ shows (Fig. 1b), in
turn, a strong drop of the conductivity on decreasing
temperature, which can be described by $\sigma \propto
\exp(-(T_o/T)^{\gamma})$ with $\gamma = 0.27 \pm 0.05$.  This
dependence points to a variable-range hopping conductivity, and
thus shows that the sample is on the insulator side of the MIT.
Here,  $ 1/4\le \gamma \le 1/2$ is expected for 3D systems,
depending on the shape of the Coulomb gap. \cite{Efro75} The
temperature dependence of the conductivity in the absence of a
magnetic field proves, therefore, the presence of the
doping--induced MIT in the Si/SiGe:Sb SL system. Moreover, the
observed increase of the conductivity at the lowest temperature
demonstrates that in the highest doped sample the
anti-localisation interaction terms dominate over other quantum
corrections to the conductivity.

Results of detailed low-temperature conductivity studies of the
$\tilde{N} = 4.5$ SL up to magnetic fields of 27 T are presented
in Fig. 2.  The orientation of the magnetic field was chosen to
be either parallel (Fig. 2a) or perpendicular (Fig. 2b) to the
SL layers.  In both cases the measurement current was
perpendicular to the magnetic field.  In the case of
sufficiently strong parallel magnetic fields, the in--plane
conductivity (inverse in--plane resistivity) $\sigma$
extrapolated to zero temperature vanishes in the $T^{1/2}$
presentation of Fig. 2a, and thus shows a magnetic-field
induced MIT.  For $\sigma$, a $T^{1/2}$
dependence, characteristic for 3D systems in the magnetic field,
\cite{Rose83,Stup93,Long84,Dai93} is visible over a wide
temperature and field range.  However, because the quantum
corrections to the conductivity of anisotropic systems are
proportional to the ratio of the diffusion constant along the
current to the average one,\cite{Alts85} the magnitude of
d$\sigma$/d$T^{1/2}$ in n-Si/SiGe SLs at criticality is found to
be by about a factor of 8 greater than in bulk n-Si.
\cite{Stup93,Long84,Dai93} This makes the extrapolation to $T =
0$ unusually inaccurate in our system and precludes an accurate 
determination of the critical magnetic field.  In contrast to the
$\tilde{N} = 4.5$ sample, the $\tilde{N} = 6.0$ sample does not
show a MIT in the magnetic field.  The extrapolated zero-$T$
conductivity $\sigma$ decreases gradually with the in-plane
magnetic field but remains non-zero up to 27 T.

{\fd}

In the case of perpendicular geometry (Fig.~2b) the decrease in
conductivity with increasing magnetic field is much weaker than
for the parallel orientation, so that no MIT is observed.  The
extrapolated zero-$T$ value of $\sigma$ does not vanish up to 12
T for the $\tilde{N} = 4.5$ SL.  For higher magnetic fields,
however, the conductivity does not decrease monotonically any
more. Accordingly, this region is not displayed in Fig. 2b as
the overlapping curves would not give a clear picture.  As shown
in the inset of Fig.~2b, in the range of high perpendicular
magnetic fields, the resistivity exhibits a broad minimum around
21 T which is consistent with a Shubnikov-de Haas oscillation or
a quantum Hall state at a filling factor $\nu = 4$.  The
occurrence of quantum Hall states is possible in the case of
weakly coupled wells in a superlattice structure.
\cite{Stor86,Chal95} The existence of quantum Hall plateaux
could not, however, be proved in our structures, as the Hall voltage is
much smaller than the longitudinal voltage in these low mobility
samples. 

We note that the occurrence of the transition for $B$ {\em parallel} 
to the interface planes (Fig.~2a), and its absence in the 
perpendicular configuration (Fig.~2b) proves rather directly that the 
field--induced MIT in n-Si \cite{Stup93,Long84,Dai93} is driven by a 
spin mechanism, not by an orbital effect.  The spin mechanism operates 
at any orientation of the magnetic field but, according to our studies 
in the weakly localised regime,\cite{Brun96} its contribution in the 
perpendicular geometry is partly compensated by an orbital effect.  
The latter was assigned\cite{Brun96} to the destructive influence of 
the perpendicular component of the magnetic field upon the 
single-electron interference of self-crossing 
trajectories.\cite{Alts85,Been91} Owing to the low mobility of the 
electrons in our structures, the interference effect compensate the 
positive magnetoresistance over a wider field range in Si/SiGe: Sb SLs 
than in high mobility Si-MOSFETs.  It should be noted however that the 
actual nature of the metallic ground state in Si-MOSFETs is unknown.

It is worth comparing the behaviour of Si-based structures to 
materials with a much larger ratio of the cyclotron to Zeeman energy, 
such as n-GaAs.  The effect of the field on the single-electron 
interference was shown to drive barely insulating n-GaAlAs to the 
metallic phase.\cite{Kats87} In sufficiently strong fields, in turn, a 
field-induced localisation is observed, which in the metallic 
n-GaAs/AlGaAs SLs occurs at smaller magnetic fields when the field is 
oriented {\em perpendicular} to the interface.\cite{Hilb95} This 
suggests that an orbital effect, such as the diamagnetic lowering of 
the carrier kinetic energy by the magnetic field, \cite{Ohts86} 
accounts for the MIT in n-GaAs-based systems.

A least-square fit to all data for the $\tilde{N} = 4.5$ sample
between 100 and 500 mK in both parallel and perpendicular
configurations was performed by the dependence $\delta\sigma(T)
\propto T^{\alpha}$.  As shown in Fig. 3, the exponent
$\alpha$ is $\approx 0.2$ at zero or weak magnetic fields, which
can be compared with $\alpha = 1/3$, as found in a recent study of n-Ge
near the MIT. \cite{Shli96}  Between 0.5 and 1.5 T a nearly step
like increase to $\alpha \approx 0.5$ takes place. In this
region the spin splitting $g\mu_{B}B$ becomes greater than the
energy of thermal broadening $kT$, and thus a cross-over to a
different universality class is expected. \cite{Beli94,Fink84} A
change of the universality class is also confirmed \cite{Brun96}
by replotting the experimental results between 0.8 and 1.4 T in
coordinates suggested by the Finkelstein renormalization group
approach \cite{Beli94,Fink84} for the MIT driven by the
disorder-modified electron-electron interactions. Because of a
strong dependence of $\sigma$ on $T$ our data are particularly
sensitive to the value of the dynamic exponent $z$, and give $z
= 1.6 \pm 0.2$ in the magnetic field.  A similar analysis for
bulk n-Si at $B=0$ yielded $z=3.1$, \cite{Stup94} while that for
the MIT of spin-polarised electrons in diluted magnetic
semiconductors resulted in $z = 1.7 \pm 0.4$. \cite{Wojt86} 

{\ft}

The values of $\alpha$ and $z$ quoted above indicate that n-Si
and n-Ge belong to the same universality class whereas diluted
magnetic semiconductors \cite{Wojt86} and n-Si in the magnetic
field \cite{Dai93} form another one.  Since the electrons in the
minority-spin band become localised and decoupled from from the
majority-spin carriers already on the metal side of the MIT, we
suggest that the corresponding universality class is rather
"spin polarised" \cite{Wojt86,Fink90} than "strong magnetic
field".\cite{Beli94,Fink84,Fink90}

In conclusion, the magnetic field-induced metal-insulator
transition in Si/SiGe superlattice structures doped with Sb is
observed for the magnetic field that is parallel to the SL
interfaces. The absence of the transition in the case of the
perpendicular magnetic fields shows rather directly that spin
effects are crucial. The temperature dependence of the
conductivity at criticality, together with detailed studies of
the magnetoconductivity in the weakly localised regime,
\cite{Brun96} demonstrate that the dominant contribution comes
from the anti-localising triplet diffusion term of the
disorder-modified electron-electron interactions.  We suggest
that this term, perhaps together with the singlet cooperon
contribution, plays also an important role in the high mobility
Si-MOS structures, where an unexpected 2D-metallic state has
been observed and shown to be destroyed by the parallel magnetic
fields.

\section*{Acknowledgements}
During the work with the dilution refrigerator on the hybrid
magnet, P. van den Linden gave exceptional technical
assistance.  This work was supported by FWF Vienna, Austria,
Polish KBN (Project No. 2-P03B-6411), and Austrian Ost--West
Program.

\frenchspacing
\vskip-12pt


\end{document}